\documentclass[12pt]{article}
\newcommand\be{\begin{equation}}
\newcommand\ee{\end{equation}}
\newcommand\bea{\begin{eqnarray}}
\newcommand\eea{\end{eqnarray}}
\newcommand\ket[1]{|#1\rangle}

\newcommand\braket[2]{\langle #1|#2\rangle}
\newcommand{\fatalpha}{{\bf \alpha \kern -0.44em \alpha}}
\newcommand{\fatsigma}{{\bf \sigma \kern -0.54em \sigma}}
\newcommand{\tpchi}{{\bf \chi \kern -0.35em \chi}}
\newcommand{\llambda}{{\bf \lambda \kern -0.45em \lambda}}

\bibliography{plain}
\title{\bf Quantum state transfer on distance regular spin networks with intrinsic decoherence}\vspace{20mm}
\author{ R. Sufiani$^{a}$
  \thanks{E-mail:sofiani@tabrizu.ac.ir} and
 A. Pedram
 \thanks{E-mail:{alipedram01@gmail.com}}
 \\ $^a$ {\small Department of Theoretical Physics and Astrophysics,
University of Tabriz, Tabriz 51664, Iran.} } \pagebreak

\vspace{20mm}
\usepackage{graphicx}
\begin{document}
\maketitle \vspace{15mm}
\newpage
\begin{abstract}
  By considering distance-regular graphs as spin networks, we investigate the state transfer fidelity in this class of networks.
  The effect of environment on the dynamics of state transfer is modeled using Milburn's intrinsic decoherence
  [G. J. Milburn, Phys. Rev. A 44, 5401 (1991)]. We consider a particular type of spin Hamiltonians
  which are extended version of those of Christandl et al [Phys. Rev. A 71, 032312 (2005)]. It is shown
  that decoherence destroys perfect communication channels. Using optimal coupling strengths derived by
  Jafarizadeh and Sufiani [Phys. Rev. A 77, 022315 (2008)], we show that
  destructive effect of environment on the communication channel increases by increasing the decoherence rate,
  however the state transfer fidelity reaches a steady value as time approaches infinity which is independent of the decoherence rate. Moreover, it is shown that for a given
  decoherence rate, the fidelity of transfer decreases by increasing the distance between the sender and the receiver.\\


{\bf PACs Index: 03.65.Ud }
\end{abstract}
\vspace{70mm}
\newpage
\section{Introduction}
The transfer of a quantum state from one part of a physical unit,
e.g., a qubit, to another part is a crucial ingredient for many
quantum information processing (QIP) protocols \cite{1}. In a
quantum-communication protocol, the transfer of quantum data from
one location $A$ to another one $B$, can be achieved by spin
networks with engineered Hamiltonians for perfect state transfer
(PST) and suitable coupling strengths between spins(see for instance
\cite{pst} and \cite{Bose}-\cite{kostak}). Particularly, Christandl,
et. al showed that PST over long distances can be implemented by a
modulated $N$-qubit XX chain (with engineered couplings between
neighborhood spins). Then, Jafarizadeh and Sufiani in \cite{pst}
extended the Christandl's work to any arbitrary distance-regular
graph as spin network and showed that by choosing suitable couplings
between spins, one can achieve to the unit fidelity of state
transfer.  Apart from these works, it could be noticed that real
systems (open systems) suffer from unavoidable interactions with
their environment (outside world). These unwanted interactions show
up as noise in quantum information processing systems and one needs
to understand and control such noise processes in order to build
useful quantum information processing systems. Open quantum systems
occur in a wide range of disciplines, and many tools can be employed
in their study. The dynamics of open quantum systems have been
studied extensively in the field of quantum optics. The main
objective in this context is to describe the time evolution of an
open system with a differential equation which properly describes
non-unitary behavior. This description is provided by the master
equation, which can be written most generally in the Lindblad form.
In fact, the master equation approach describes quantum noise in
continuous time using differential equations, and is the approach to
quantum noise most often used by physicists.

In the state transfer scenarios , the decoherence effects of the
system-environment interactions avoid us to achieve unit fidelity.
Study of the influence of different kinds of noise on the fidelity
of quantum state transfer has seldom considered so far. Recently, M.
L. Hu, et al \cite{hu} have studied state transfer over an $N$-qubit
open spin chain with intrinsic decoherence and a chin with dephasing
environment\cite{hu2}. By solving the corresponding master equation
analyticaly, they investigated optimal state transfer and also
creation and distribution of entanglement in the model of Milburn's
intrinsic decoherence. In this work, we extend their approach to any
arbitrary distance regular spin network (DRSN) in order to transfer quantum
data with optimal fidelity over the antipodes of these networks. In
fact it is shown that, the optimal transfer fidelity depends on the
spectral properties of the networks, and the desired fidelity is
evaluated in terms of the polynomials associated with the networks.
Moreover, a closed formula for the steady fidelity $F^{(s)}(t)$ at
large enough times $t$ is given in terms of the corresponding
polynomials.

The organization of the paper is as follows. In section 2, some
preliminaries such as definition of distance regular networks,
technique of stratification and spectral distribution for these
networks are reviewed. In section $3$, the model describing
interactions of spins with each others- via a distance regular
network- is introduced and the Milburn's intrinsic decoherence is
employed in order to obtain the optimal fidelity of state transfer
as the main result of the paper. Section $4$ is concerned with some
important examples of distance regular networks where the fidelity
of transfer is evaluated in each case. Paper is ended with a brief
conclusion.
\section{Preliminaries}
\subsection{Distance-regular networks and stratification}
Distance-regular graphs lie in an important category of graphs which
possess some useful properties. In these graphs, the adjacency
matrices $A_i$ are defined based on shortest path distance denoted
by $\partial$. More clearly, if $\partial(\alpha, \beta)$ (distance
between the nodes $\alpha,\beta\in V$) be the length of the shortest
walk connecting $\alpha$ and $\beta$ (recall that a finite sequence
$\alpha_0, \alpha_1,..., \alpha_n \in V$ is called a walk of length
$n$ if $\alpha_{k-1}\sim \alpha_k$ for all $k=1, 2,..., n$, where
$\alpha_{k-1}\sim \alpha_k$ means that $\alpha_{k-1}$ is adjacent
with $\alpha_{k}$), then the adjacency matrices $A_i$ for
$i=0,1,...,d$ in distance-regular graphs are defined as:
$(A_i)_{\alpha,\beta}=1$ if and only if $\partial(\alpha, \beta)=i$
and $(A_i)_{\alpha,\beta}=0$ otherwise, where
$d:=$max$\{\partial(\alpha, \beta): \alpha, \beta\in V \}$ is called
the diameter of the graph. In fact, an undirected connected graph
$G=(V,E)$ is called distance-regular graph (DRG) with diameter $d$
if for all $i,j,k\in\{0,1,...,d\}$, and $\alpha,\beta$ with
$\partial(\alpha,\beta)=k$, the number
\begin{equation}\label{def}
p^k_{ij}=\mid \{\gamma\in V : \partial(\alpha, \gamma)=i \;\ and \;\
\partial(\gamma,\beta)=j\}\mid
\end{equation}
is constant in that it depends only on $k, i, j$ but does not depend
on the choice of $\alpha$ and $\beta$. Some more details about these
graphs have been given in the section $B. I$ of the Appendix $B$ of
Ref.\cite{pst}.

For a given vertex $\alpha\in V$, let $\Gamma_i(\alpha):=\{\beta\in
V: \partial(\alpha, \beta)=i\}$ denotes the set of all vertices
being at distance $i$ from $\alpha$. Then, the vertex set $V$ can be
written as disjoint union of $\Gamma_i(\alpha)$ for $i=0,1,2,...,d$,
i.e.,
\begin{equation}\label{asso1}
V=\bigcup_{i=0}^{d}\Gamma_{i}(\alpha),
\end{equation}
In fact, by fixing a point $o\in V$ as a reference vertex, the
relation (\ref{asso1}) stratifies the vertices into a disjoint union
of associate classes $\Gamma_{i}(o)$ (called the $i$-th stratum or
$i$-th column with respect to $o$). With each associate class
$\Gamma_{i}(o)$ we associate a normalized vector in $l^2(V)$ (the
Hilbert space of all square summable functions in $V$) defined by
\begin{equation}\label{unitv}
\ket{\phi_{i}}=\frac{1}{\sqrt{\kappa_i}}\sum_{\alpha\in
\Gamma_{i}(o)}\ket{\alpha},
\end{equation}
where, $\ket{\alpha}$ denotes the eigenket of $\alpha$-th vertex at
the associate class $\Gamma_{i}(o)$ and $\kappa_i=|\Gamma_{i}(o)|$
is called the $i$-th valency of the graph. The space spanned by
$\ket{\phi_i}$s, for $i=0,1,...,d$  is called ``stratification
space". Then, the adjacency matrix reduced to this space satisfies
the following three term recursion relation
\begin{equation}\label{trt}
A\ket{\phi_i}=\beta_{i+1}\ket{\phi_{i+1}}+\alpha_i\ket{\phi_i}+\beta_i\ket{\phi_{i-1}}.
\end{equation}
i.e., the adjacency matrix $A$ takes a tridiagonal form in the
orthonormal bases $\{\ket{\phi_i}, i=0,1,\ldots, d-1\}$, so that, for
spin networks of distance-regular type we can restrict our attention
to the stratification space for the purpose of state transfer from
$\ket{\phi_0}$ to $\ket{\phi_d}$ (state associated with the last
stratum of the network). For the purpose of optimal state transfer,
we will deal with particular distance-regular graphs (as spin
networks) for which starting from an arbitrary vertex as reference
vertex (prepared in the initial state which we wish to transfer),
the last stratum of the networks with respect to the reference state
contains only one vertex.

Now, let $A_i$ be the $i$th adjacency matrix of the graph
$\Gamma=(V,E)$. Then, for the reference state $\ket{\phi_0}=\ket{o}$
one can write
\begin{equation}\label{Foc1}
A_i\ket{\phi_0}=\sum_{\beta\in \Gamma_{i}(o)}\ket{\beta}.
\end{equation}
Then, by using (\ref{unitv}) and (\ref{Foc1}), we obtain
\begin{equation}\label{Foc2}
A_i\ket{\phi_0}=\sqrt{\kappa_i}\ket{\phi_i}.
\end{equation}
It can be shown that \cite{pst}, for the adjacency matrices $A_i$ of
distence regular graphs, we have
\begin{equation}\label{P1}
A_i=P_i(A),\;\ i=0,1,...,d,
\end{equation}
where, $P_i$ is a polynomial of degree $i$. Then, the
Eq.(\ref{Foc2}) gives
\begin{equation}\label{fi}
\ket{\phi_i}=\frac{P_i(A)}{\sqrt{\kappa_i}}\ket{\phi_0}.
\end{equation}
\subsection{Spectral techniques}
In this subsection, we recall some preliminary facts about spectral
techniques used in the paper, where more details have been given in
the appendix $B. II$ of \cite{pst} and Refs. \cite{js}-\cite{js1}.

Actually the spectral analysis of operators is an important issue in
quantum mechanics, operator theory and mathematical physics
\cite{simon, Hislop}. As an example $\mu(dx)=|\psi(x)|^2dx$
($\mu(dp)=|\widetilde{\psi}(p)|^2dp$) is a spectral distribution
which is  assigned to  the position (momentum) operator
$\hat{X}(\hat{P})$.

It is well known that, for any pair $(A,\ket{\phi_0})$ of a matrix
$A$ and a vector $\ket{\phi_0}$, one can assign a measure $\mu$ as
follows
\begin{equation}\label{sp1}
\mu(x)=\braket{ \phi_0}{E(x)|\phi_0},
\end{equation}
 where
$E(x)=\sum_i|u_i\rangle\langle u_i|$ is the operator of projection
onto the eigenspace of $A$ corresponding to eigenvalue $x$, i.e.,
\begin{equation}
A=\int x E(x)dx.
\end{equation}
Then, for any polynomial $P(A)$ we have
\begin{equation}\label{sp2}
P(A)=\int P(x)E(x)dx,
\end{equation}
where for discrete spectrum the above integrals are replaced by
summation. Therefore, using the relations (\ref{sp1}) and
(\ref{sp2}), the expectation value of powers of adjacency matrix $A$
over reference vector $\ket{\phi_0}$ can be written as
\begin{equation}\label{v2}
\braket{\phi_{0}}{A^m|\phi_0}=\int_{R}x^m\mu(dx), \;\;\;\;\
m=0,1,2,....
\end{equation}
Obviously, the relation (\ref{v2}) implies an isomorphism from the
Hilbert space of the stratification onto the closed linear span of
the orthogonal polynomials with respect to the measure $\mu$.
Moreover, using the correspondence $A\equiv x$ and the equations
(\ref{trt}) and (\ref{fi}), one gets three term recursion relations
between polynomials $P_i(x)$
\begin{equation}\label{eq6}
\beta_{i+1}\frac{P_{i+1}(x)}{\sqrt{\kappa_{i+1}}}=(x-\alpha_i)\frac{P_{i}(x)}{\sqrt{\kappa_{i}}}-\beta_i\frac{P_{i-1}(x)}{\sqrt{\kappa_{i-1}}}
\end{equation}
for $i=0,...,d-1$. Multiplying by $\beta_1...\beta_i$ we obtain
\begin{equation}
\beta_1...\beta_{i+1}\frac{P_{i+1}(x)}{\sqrt{\kappa_{i+1}}}=(x-\alpha_i)\beta_1...\beta_i\frac{P_{i}(x)}{\sqrt{\kappa_{i}}}-\beta_i^2.\beta_1...\beta_{i-1}\frac{P_{i-1}(x)}{\sqrt{\kappa_{i-1}}}.
\end{equation}
By rescaling $P_i$ as
$Q_i=\beta_1...\beta_i\frac{P_i}{\sqrt{\kappa_i}}$, the three term
recursion relations (\ref{eq6}) are replaced by
$$ Q_0(x)=1, \;\;\;\;\;\
Q_1(x)=x,$$
\begin{equation}\label{op}
Q_{i+1}(x)= (x-\alpha_{i})Q_i(x)-\beta_i^2Q_{i-1}(x),
\end{equation}
for $i=1,2, ..., d$.

In the next section, we will need the distinct eigenvalues of
adjacency matrix of a given undirected graph and the corresponding
eigenvectors in order to obtain the evolved density matrix and the
corresponding fidelity of state transfer. As it is known from
spectral theory, we have the eigenvalues $x_k$ of the adjacency
matrix $A$ as roots of the last polynomial $Q_{d+1}(x)$ in
(\ref{op}), and the normalized eigenvectors as \cite{tsc, lipkin}
\begin{equation}\label{eigvec}\ket{\psi_k}=\frac{1}{\sqrt{\sum_{l=0}^dP^2_l(x_k)}}\left(\begin{array}{c}
                                                          P_0(x_k) \\
                                                              P_1(x_k) \\
                                                              \vdots \\
                                                              P_d(x_k) \\
                                                            \end{array}\right),
                                                            \end{equation}

                                                            in which
                                                            we have
                                                            $P_i(x)={\frac{\sqrt{\kappa_i}}{\beta_1\ldots
                                                            \beta_i}}Q_i(x)$ for $i=0,1,\ldots,d$.
\section{The model}
The model we will consider is a distance-regular network consisting
of $N$ sites labeled by $\{1,2, ... ,N\}$ and diameter $d$. Then we
stratify the network with respect to a chosen reference site, say
$1$, and assume that the network  contains only the output site $N$
in its last stratum (i.e., $\ket{\phi_d}=\ket{N}$). At time $t=0$,
the qubit in the first (input) site of the network is prepared in
the state $\ket{\psi_{in}}$. We wish to transfer the state to the
$N$th (output) site of the network with unit efficiency after a
well-defined period of time. Although our qubits represent generic
two state systems, for the convenience of exposition we will use the
term spin as it provides a simple physical picture of the network.
The standard basis for an individual qubit is chosen to be $\{
\ket{0}=\ket{\downarrow},\;\ \ket{1}=\ket{\uparrow} \}$, and we
shall assume that initially all spins point ``down" along a
prescribed $z$ axis; i.e., the network is in the state
$\ket{\b{0}}=\ket{0_A00...00_B}$. Then, we consider the dynamics of
the system to be governed by the quantum-mechanical Hamiltonian
\begin{equation}\label{H}
H_G =\frac{1}{2} \sum_{m=0}^dJ_{m}\sum_{_{(i,j)\in R_m}}H_{ij},
\end{equation}
with $H_{ij}$ as
\begin{equation}
H_{ij} = {\mathbf{\sigma}}_i\cdot {\mathbf{\sigma}}_j,
\end{equation}
where, ${\mathbf{\sigma}}_i$ is a vector with familiar Pauli
matrices $\sigma^x_i, \sigma^y_i$ and $\sigma^z_i$  as its
components acting on the one-site Hilbert space ${\mathcal{H}}_i$,
and $J_{m}$ is the coupling strength between the reference site $1$
and all of the sites belonging to the $m$-th stratum with respect to
$1$.

Then, by employing the symmetry corresponding to conservation of $z$
component of the total spin and reduction of the main Hilbert space
to the single excitation subspace spanned by the basis $\{ \ket{l},
l=1,\ldots, N \}$ with $\ket{l}=\ket{0 ... 0 \underbrace{1}_{l-th} 0
... 0}$, the hamiltonian in (\ref{H}) can be written in terms of the
adjacency matrices $A_i$, $i=0,1,...,d$ as follows
\begin{equation}\label{HA}
H=2\sum_{m=0}^dJ_mA_m+\frac{N-4}{2}\sum_{m=0}^dJ_m\kappa_m I.
\end{equation}
For details, see \cite{pst}. For the purpose of the optimal transfer
of state, as we will see in the next subsection, we will need only
the difference between the energies, and so the second constant term
in (\ref{HA}) dose not affect the last desired result and can be
neglected for simplicity.

Now, using the Eq.(\ref{P1}) and restricting the $N$-dimensional
single excitation subspace to the $(d+1)$-dimensional Krylov
subspace spanned by the stratification basis $\{\ket{\phi_0},
\ket{\phi_1}, \ldots, \ket{\phi_{d-1}}\}$, we can write
\begin{equation}\label{HA'}
H=2\sum_{m=0}^dJ_mP_m(A)=2\sum_{m=0}^dJ_m\sum_{k=0}^dP_m(x_k)\ket{\psi_k}\langle
\psi_{k'} |,
\end{equation}
where $x_k$ and $\ket{\psi_k}$ for $k=0,1,\ldots, d$ are the
corresponding eigenvalues and eigenvectors of the adjacency matrix
$A$ given by the spectral method illustrated in the previous
sections (given by Eq.(\ref{eigvec})). Obviously, the energy
eigenvalues are given by
\begin{equation}\label{eigenvalue}
E_k=2\sum_{m=0}^dJ_mP_m(x_k).
\end{equation}
\subsection{Milburn's intrinsic decoherence}
Milburn in ref. \cite{milburn} has assumed that on sufficiently
short time steps, the system does not evolve continuously under
unitary evolution but rather in a stochastic sequence of identical
unitary transformations which can account for the disappearance of
quantum coherence as the system evolves. With this assumption,
Milburn obtained the master equation for the evolution of the system
as
\begin{equation}\label{eq1}
\frac{d\rho}{d t}=\frac{1}{\gamma} \{\exp (-i\gamma H)\rho \exp
(i\gamma H)-\rho\},
\end{equation}
where $\gamma$ is the intrinsic decoherence parameter and $H$ is the
Hamiltonian of the considered system. Expanding the above equation,
keeping terms up to first order in $\gamma$ gives
\begin{equation}\label{eq2}
\frac{d\rho}{d t}=-i[H,\rho]-\frac{\gamma}{2} [H,[H,\rho]].
\end{equation}
The second term on the r.h.s of the above equation represents the
decoherence effect on the system, where in the limit of
$\gamma\rightarrow 0$ the ordinary Schr\"{o}dinger equation is
recovered. By defining three auxiliary super-operators $J$, $S$ and
$L$ as
$$J\rho=\gamma H\rho H,\;\;\ S\rho=-i[H,\rho],\;\;\ L\rho=-\frac{\gamma}{2}\{H^2,\rho\},$$
one can straightforwardly show that \cite{hu} the Eq.(\ref{eq2})
simplifies to ${\frac{d\rho}{dt}}=(J+S+L)\rho$, where its solution
can be written in terms of the so called Kraus operators $K_l(t)$ as
\begin{equation}\label{eq3}
\rho(t)=\sum^{{\infty}}_{_{l=0}}K_l(t)\rho(0)K^{{\dag}}_l(t).
\end{equation}
In the Eq.(\ref{eq3}), $\rho(0)$ denotes the initial state of the
system and the Kraus operators $K_l(t)=(\gamma t)^{l/2}H^l\exp
(-iHt)\exp (-\gamma tH^2/2)/\sqrt{l!}$ satisfy the relation
$\sum^{{\infty}}_{l=0} K^{{\dag}}_l(t)K_l(t)=I$ for all times $t$.

Now by expanding $\rho(0)$ in terms of energy eigenstates as
$\rho(0)=\sum_{k,k'}a_{kk'}\ket{\psi_k}\langle {\psi_{k'}}|$, one
can obtain
\begin{equation}\label{eq4}
\rho(t)=\sum_{k,k'}a_{kk'}\exp[-it(E_k-E_{k'})-\frac{\gamma
t}{2}(E_k-E_{k'})^2]|\psi_k\rangle\langle \psi_{k'}|
\end{equation}
where $a_{kk'}=\langle \psi_k|\rho(0)\ket{\psi_{k'}}$, $E_k$ and
$\ket{\psi_k}$ are eigenvalues and the corresponding eigenvectors of
the system. Consider the initial state as
$$
\rho(0)=|{\phi_0}\rangle\langle {\phi_0}|=\sum a_{kk'}
|{\psi_k}\rangle\langle {\psi_{k'}}|,
$$
where, using Eq.(\ref{eigvec}), we have
$$a_{kk'}=\langle {\psi_k}|{\phi_0}\rangle \langle {\phi_0}|{\psi_{k'}}\rangle=\frac{P_0(x_k)P_0(x_{k'})}{\sqrt{\sum_l {P_l}^2(x_k)\sum_l {P_l}^2(x_{k'})}}=\frac{1}{\sqrt{\sum_l {P_l}^2(x_k)\sum_l {P_l}^2(x_{k'})}}.$$
Then, Eq.(\ref{eq4}) gives us
\begin{equation}\label{rot}
\rho(t)=\sum_{k,k'}{\frac{\exp{[-it(E_k-E_{k'})-\frac{\gamma
t}{2}(E_k-E_{k'})^2]}}{\sqrt{\sum_l {P_l}^2(x_k)\sum_l
{P_l}^2(x_{k'})}}}|\psi_k\rangle\langle \psi_{k'}|.
\end{equation}
Now, employing the Eq.(\ref{eigvec}), the fidelity of state transfer
is given by
\begin{equation}\label{fidelity}
F(t)=\langle
{\phi_d}|\rho(t)|{\phi_d}\rangle=\sum_{k,k'}\frac{\exp[-it(E_k-E_{k'})-\frac{\gamma
t}{2}(E_k-E_{k'})^2]}{{\sum_l {P_l}^2(x_k)\sum_l {P_l}^2(x_{k'})}}
P_d(x_k)P_d(x_{k'}),
\end{equation}
where the eigenvalues $E_k$ are evaluated via the Eq.
(\ref{eigenvalue}). On the other hands, one can show that for
distance regular networks, we have $P_d(x_k)=(-1)^k$ (see Refs.
\cite{Godsil}, \cite{brower}). Therefore the fidelity
(\ref{fidelity}) is simplified to
\begin{equation}\label{fidelity''}
F(t)=\langle
{\phi_d}|\rho(t)|{\phi_d}\rangle=\sum_{k,k'}(-1)^{k+k'}\frac{\exp[-it(E_k-E_{k'})-\frac{\gamma
t}{2}(E_k-E_{k'})^2]}{{\sum_l {P_l}^2(x_k)\sum_l {P_l}^2(x_{k'})}}.
\end{equation}
Due to the exponential term in the Eq.(\ref{fidelity}), for large
enough times $t\rightarrow {\infty}$ the transfer fidelity tends to
zero except for $k$,$k'$ with $E_k=E_{k'}$.  Therefore, after large
enough times $t\gg$, the transfer fidelity arrives at a stable value
as follows
\begin{equation}\label{fidelity1}
F^s(t\rightarrow\infty)=\sum_{k=0}^{d}\frac{1}{[\sum_l
{P_l}^2(x_{k})]^2}+2\sum_{{k<k'}_{with {
E_k=E_{k'}}}}^{d}\frac{(-1)^{k+k'}}{\sum_l {P_l}^2(x_{k})\sum_l
{P_l}^2(x_{k'})}.
\end{equation}
In fact for the networks for which $E_k\neq E_{k'}$ for $k\neq k'$,
one obtains the steady fidelity as
\begin{equation}\label{fidelity1'}
F^s(t\rightarrow\infty)=\sum_{k=0}^{d}\frac{1}{[\sum_l
{P_l}^2(x_{k})]^2}=\sum_{k=0}^{d}a^2_{kk}.
\end{equation}
\section{Examples}
\subsection{The Cyclic network with even number of nodes}
A well known example of distance-regular networks, is the cycle
graph with $N$ vertices denoted by $C_N$. For the purpose of optimal
state transfer, we consider the even number of vertices $N=2m$ for
which the last stratum contains a single state corresponding to the
$m$-th node. The adjacency matrices are given by
\begin{equation}\label{adjcyce.}
A_0=I_{2m},\;\;\ A_i=S^i+S^{-i},\;\ i=1,2,...,m-1, \;\;\ A_m=S^m,
\end{equation}
where, $S$ is the $N\times N$ circulant matrix with period $N$ (i.e.
$S^N=I_N$). The corresponding parameters $\alpha_i$ and $\beta_i$
are given by
\begin{equation}\label{QDcye.}
\alpha_i=0, \;\ i=0,1,...,m; \;\ \beta_1=\beta_m=\sqrt{2},\;\
\beta_i=1,\;\ i=2,...,m-1.
\end{equation}
Now, using the recursion relations (\ref{op}), one can show that \be
Q_0(x)=P_0(x)=1,\;\ Q_i(x)=P_i(x)=2T_i(x/2) ,\;\ i=1,..., m-1,\;\;\
Q_m(x)=2P_m(x)=2T_m(x/2),\ee where $T_i$'s are Tchebychev
polynomials of the first kind. Then, the eigenvalues of the
adjacency matrix $A\equiv A_1$ are given by
$$x_i=\omega^i+\omega^{-i}=2\cos(2\pi i/N)=2\cos(\pi i/m),\;\ i=0,1,...,m$$
with $\omega:=e^{2\pi i/N}$. Details have given in Ref.\cite{pst}.

By choosing the suitable coupling constants given in \cite{1}, and
Eq. (\ref{fidelity''}), the fidelity of transfer can be plotted. In
Figure 1, the fidelity of transfer has plotted for $N=4$ with three
different decoherence rates $\gamma=0.1$, $\gamma=0.2$ and
$\gamma=0.3$ and coupling constants $J_0=-J_2=-\pi/4$ and $J_1=0$.
Figure 2 shows the fidelity of transfer for the cases $N=4$, $N=6$ and
$N=8$ with $\gamma=0.1$ and the same coupling strengths. As it is
seen from Figure 1 and Figure 2, the fidelity decreases with increasing
decoherence rate $\gamma$; and for a given $\gamma$, the optimal
probability of transfer decreases by distance.
\begin{figure}[h!]
  \centering
  \hspace{-4cm}\includegraphics[totalheight=8cm]{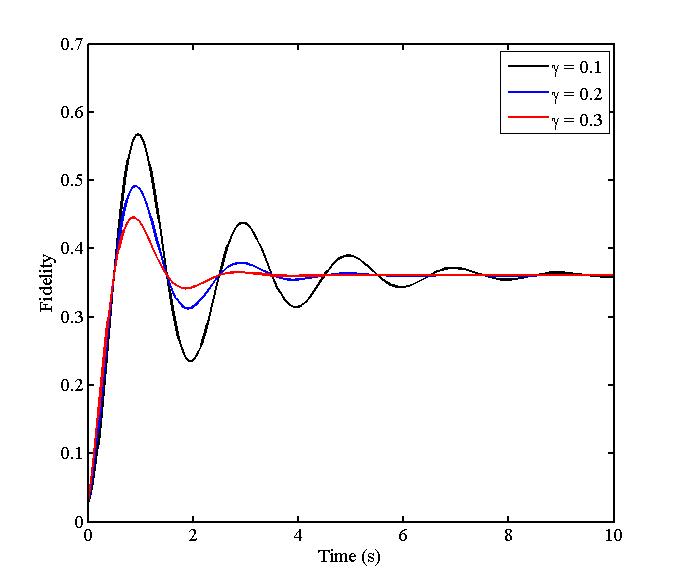}
  \caption{(Color online) Dynamics of the state transfer fidelity for $C_4$ with three different decoherence rates}\label{Fig.1}
\end{figure}

\begin{figure}[h!]
  \centering
  \hspace{-4cm}\includegraphics[totalheight=8cm]{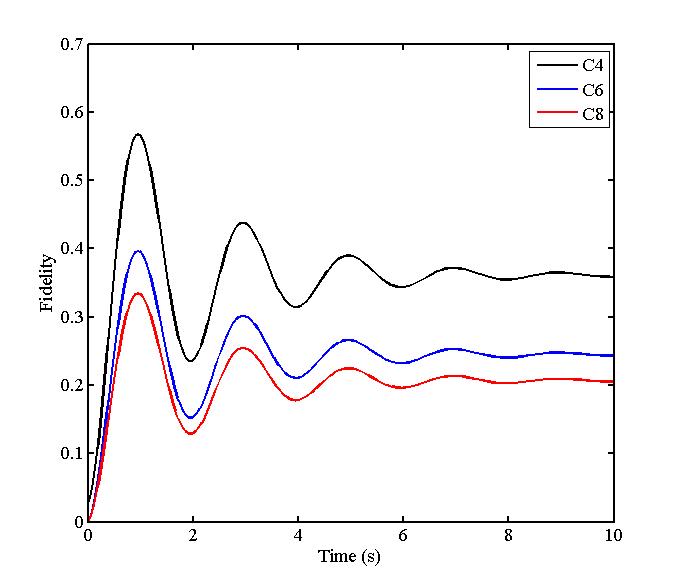}
  \caption{(Color online) Dynamics of the state transfer fidelity for $C_N$ with $N=4,6,8$ and $\gamma=0.1$}\label{Fig.2}
\end{figure}
\newpage
\subsection{The hypercube network}
The hypercube of dimension $d$ (known also as binary Hamming scheme
$H(d,2)$) is a network with $N=2^d$ nodes, each of which can be
labeled by an $d$-bit binary string. Two nodes on the hypercube
described by bitstrings $\vec{x}$ and $\vec{y}$ are are connected by
an edge if $|\vec{x}- \vec{y}|=1$, where $|\vec{x}|$ is the Hamming
weight of $\vec{x}$.

 In other words, if $\vec{x}$ and $\vec{y}$
differ by only a single bit flip, then the two corresponding nodes
on the network are connected. Thus, each of $2^d$ nodes on the
hypercube has degree $d$. For the hypercube network with dimension
$d$ we have $d+1$ strata with adjacency matrices
\begin{equation}
A_i=\sum_{perm.}\underbrace{\sigma_x\otimes\sigma_x...\otimes\sigma_x}_{i}
\underbrace{\otimes I_2\otimes...\otimes I_2}_{d-i},\;\ i=0,1,...,d,
\end{equation}
where, the summation is taken over all possible nontrivial
permutations. The eigenvalues $x_i$ and the corresponding parameters
$\alpha_i$ and $\beta_i$ for $i=0,1,...,d$, are given by $x_i=2i-d$
and
$$\alpha_i=0,\;\;\ \beta_i=\sqrt{i(d-i+1)},$$
respectively. For details see Ref.\cite{pst}. Figure 3 shows the
optimal fidelity of transfer for the case $d=3$(the known cube
network) with decoherence rates $\gamma=0.1$, $\gamma=0.2$ and
$\gamma=0.3$. We have chosen the optimal set of coupling strengths $
J_0=-\frac{3\pi}{4},\;\ J_1=\frac{\pi}{4},\;\ J_2=J_3=0$ given in
Ref.\cite{pst}. In Figure 4, the results for the cases $d=3$, $d=4$ and
$d=5$ is shown, where it is seen that the fidelity decreases by
increasing the dimension $d$ of the hypercube $H(d,2)$.
\begin{figure}[h!]
  \centering
  \hspace{-4cm}\includegraphics[totalheight=8cm]{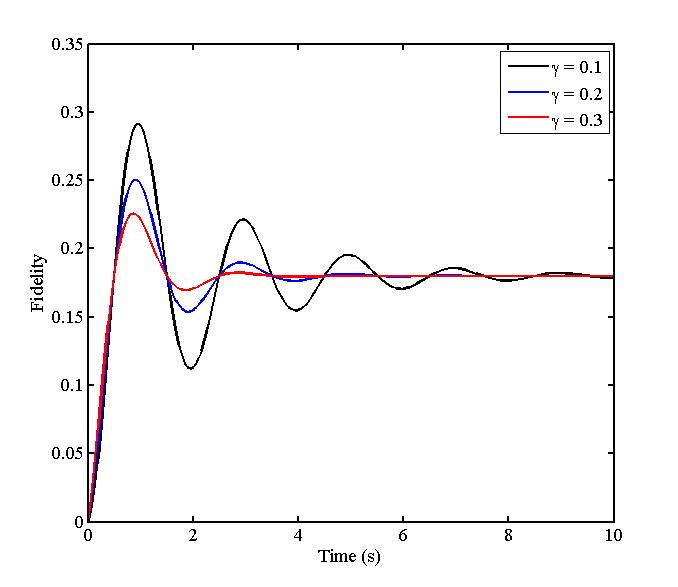}
  \caption{(Color online) Dynamics of the state transfer fidelity for the cube network $H(3,2)$ with three different decoherence rates}\label{Fig.3}
\end{figure}
\begin{figure}[h!]
  \centering
  \hspace{-4cm}\includegraphics[totalheight=8cm]{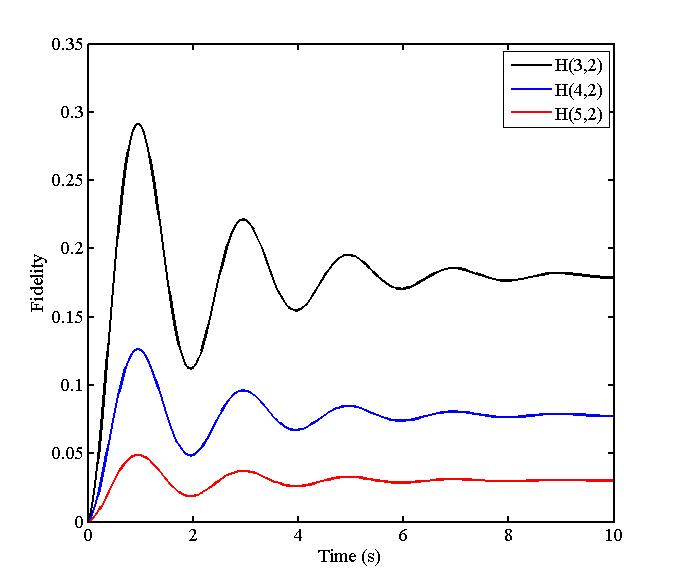}
  \caption{(Color online) Dynamics of the state transfer fidelity for $H(d,2)$ with $d=3,4,5$ and $\gamma=0.1$}\label{Fig.4}
\end{figure}
\newpage
\subsection{The Crown network}
A crown graph on $2m$ vertices is an undirected network with two
sets of vertices $u_i$ and $v_i$, where the vertex $u_i$ is
connected to $v_j$ whenever
 $i\neq j$.
 The corresponding adjacency matrix is given by $ A= K_{m}\otimes \sigma_{x}$
where, $ K_{m} $ is the adjacency matrix of the complete graph with
$m$ vertices and $ \sigma_{x}  $ is the Pauli matrix.  Then, the
stratification bases (Krylov bases) are given by
$$\hspace{-8.25cm} \ket{\phi_0}=\ket{1},$$
$$ \hspace{-1cm}\ket{\phi_1}= \frac{1}{\sqrt {n-1}}(\ket{m+1 }+\ket{m+2 } +\cdots +\ket{2m-1}),$$
$$ \hspace{-3.5cm}\ket{\phi_2}= \frac{1}{\sqrt {m-1}}(\ket{2 }+\ket{3 }+ \cdots +\ket{m}),$$
$$\hspace{-8cm} \ket{\phi_3}=\ket{2m }.$$
In the $\{\ket{\phi_i}\}$ bases, the adjacency matrix is represented
as
 $$ A=\left(
        \begin{array}{cccc}
          0 & \sqrt {m-1} & 0 & 0 \\
          \sqrt {m-1} & 0 & m-2 & 0 \\
          0 & m-2 & 0 & \sqrt {m-1} \\
          0 & 0 & \sqrt {m-1} & 0 \\
        \end{array}
      \right),
 $$
so that we have $\alpha_i=0$ for $i=0,1,2,3$ and
$\beta_1=\beta_3=\sqrt{m-1},\;\ \beta_2=m-2$. Now, by using
(\ref{op}) one can easily calculate the roots $x_i$ as follows:
$$x_0=-(m-1),\;\ x_1=-1,\;\ x_2=1, \;\ x_3=(m-1).$$
Now, following the algorithm given in \cite{pst}, one can obtain the
optimal couplings $J_i$ as follows
$$J_0=-\frac{\pi}{4},\;\;\ J_1=J_2=0,\;\;\ J_3=\frac{\pi}{4}.$$
Figure 5 shows the optimal fidelity of transfer for the case $m=3$ with
decoherence rates $\gamma=0.1$, $\gamma=0.2$ and $\gamma=0.3$. Plots
for the cases $m=3$, $m=4$ and $m=5$ with $\gamma=0.1$ are shown in
Figure 6.
\begin{figure}[h!]
  \centering
  \hspace{-4cm}\includegraphics[totalheight=8cm]{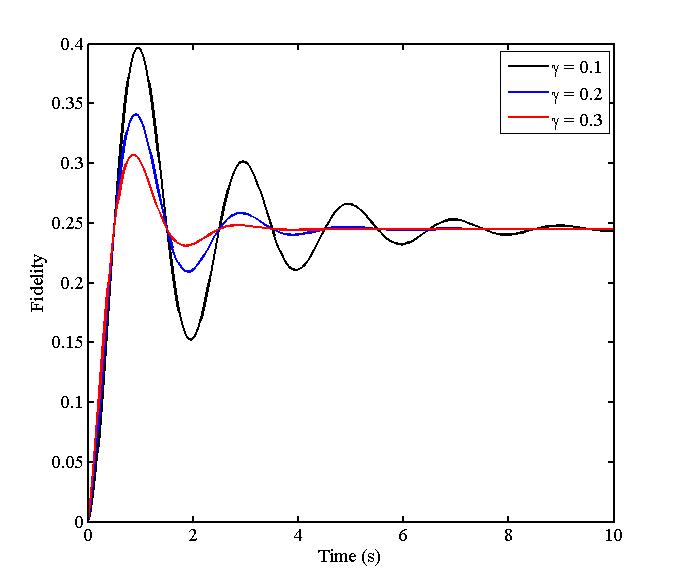}
  \caption{(Color online) Dynamics of the state transfer fidelity for the Crown network with $m=3$ and three different decoherence rates}\label{Fig.5}
\end{figure}

\begin{figure}[h!]
  \centering
  \hspace{-4cm}\includegraphics[totalheight=8cm]{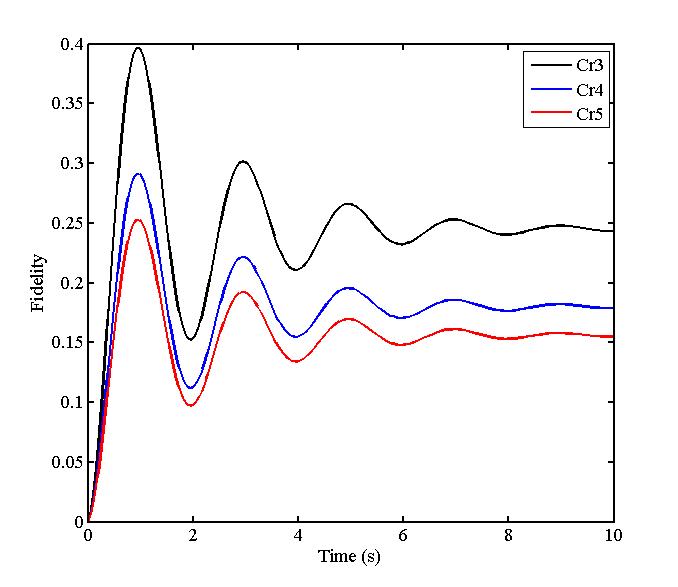}
  \caption{(Color online) Dynamics of the state transfer fidelity for the Crown network with $m=3,4,5$ and $\gamma=0.1$}\label{Fig.6}
\end{figure}
\newpage
\section{Conclusion}
In summery, optimal state transfer over distance regular spin
networks (DRSN) in the Milburn's intrinsic decoherence environment
was studied. In fact, using the spectral properties of these
networks and employing the stratification technique for them, we
obtained the transfer fidelity over DRSNs in terms of the
polynomials associated with them. By choosing the optimal coupling
constants (considered in Phys. Rev. A 77, 022315 (2008)) for perfect
state transfer (PST), it was seen that intrinsic decoherence
destroys perfect communication so that destructive effect of
environment on the communication channel increases by increasing the
decoherence rate. However the transfer fidelity reaches a steady
value as time approaches infinity which is independent of the
decoherence rate. Moreover, it was shown in some examples that for a
given decoherence rate, the fidelity of transfer decreases by
distance between the sender and the receiver (antipodes of the
corresponding networks).

\end{document}